\documentclass[aps,prl,twocolumn,superscriptaddress]{revtex4-1}
\usepackage{graphicx,longtable,graphics,amsmath}

\begin{document}
\bibliographystyle{apsrev4-1}

\title{Electronic control of the spin-wave damping in a magnetic
  insulator}

\author{A. Hamadeh}
\affiliation{Service de Physique de l'\'Etat Condens\'e (CNRS URA
  2464), CEA Saclay, 91191 Gif-sur-Yvette, France}

\author{O. d'Allivy Kelly}
\affiliation{Unit\'e Mixte de Physique CNRS/Thales and Universit\'e
  Paris Sud 11, 1 av. Fresnel, 91767 Palaiseau, France}

\author{C. Hahn}
\affiliation{Service de Physique de l'\'Etat Condens\'e (CNRS URA
  2464), CEA Saclay, 91191 Gif-sur-Yvette, France}

\author{H. Meley}
\affiliation{Service de Physique de l'\'Etat Condens\'e (CNRS URA
  2464), CEA Saclay, 91191 Gif-sur-Yvette, France}

\author{R. Bernard} 
\affiliation{Unit\'e Mixte de Physique CNRS/Thales
  and Universit\'e Paris Sud 11, 1 av. Fresnel, 91767 Palaiseau,
  France}

\author{A.H. Molpeceres} 
\affiliation{Unit\'e Mixte de Physique CNRS/Thales
  and Universit\'e Paris Sud 11, 1 av. Fresnel, 91767 Palaiseau,
  France}

\author{V. V. Naletov}
\affiliation{Service de Physique de l'\'Etat Condens\'e (CNRS URA
  2464), CEA Saclay, 91191 Gif-sur-Yvette, France}
\affiliation{Unit\'e Mixte de Physique CNRS/Thales and Universit\'e
  Paris Sud 11, 1 av. Fresnel, 91767 Palaiseau, France}
\affiliation{Institute of Physics, Kazan Federal University, Kazan
    420008, Russian Federation}

\author{M. Viret}
\affiliation{Service de Physique de l'\'Etat Condens\'e (CNRS URA
  2464), CEA Saclay, 91191 Gif-sur-Yvette, France}

\author{A. Anane}
\affiliation{Unit\'e Mixte de Physique CNRS/Thales and Universit\'e
  Paris Sud 11, 1 av. Fresnel, 91767 Palaiseau, France}

\author{V. Cros}
\affiliation{Unit\'e Mixte de Physique CNRS/Thales and Universit\'e
  Paris Sud 11, 1 av. Fresnel, 91767 Palaiseau, France}


\author{S. O. Demokritov} \affiliation{Department of Physics,
  University of Muenster, 48149 Muenster, Germany}

\author{J. L. Prieto} 
\affiliation{Instituto de Sistemas Optoelectr\'onicos y
  Microtecnolog\'{\i}a (UPM), Madrid 28040, Spain}

\author{M. Mu\~noz}
\affiliation{Instituto de Microelectr\'onica de Madrid (CNM, CSIC),
  Madrid 28760, Spain}

\author{G. de Loubens}
\affiliation{Service de Physique de l'\'Etat Condens\'e (CNRS URA
  2464), CEA Saclay, 91191 Gif-sur-Yvette, France}

\author{O. Klein}\email[Emails: ]{oklein@cea.fr \& gregoire.deloubens@cea.fr}
\affiliation{Service de Physique de l'\'Etat Condens\'e (CNRS URA
  2464), CEA Saclay, 91191 Gif-sur-Yvette, France}

\date{\today}

\begin{abstract}
  It is demonstrated that the decay time of spin-wave modes existing
  in a magnetic insulator can be reduced or enhanced by injecting an
  in-plane dc current, $I_\text{dc}$, in an adjacent normal metal with
  strong spin-orbit interaction. The demonstration rests upon the
  measurement of the ferromagnetic resonance linewidth as a function
  of $I_\text{dc}$ in a 5~$\mu$m diameter YIG(20nm){\textbar}Pt(7nm)
  disk using a magnetic resonance force microscope (MRFM). Complete
  compensation of the damping of the fundamental mode is obtained for
  a current density of $\sim 3 \cdot 10^{11}\text{A.m}^{-2}$, in
  agreement with theoretical predictions. At this critical threshold
  the MRFM detects a small change of static magnetization, a behavior
  consistent with the onset of an auto-oscillation regime.
\end{abstract}

\maketitle

The spin-orbit interaction (SOI) \cite{valenzuela06, Jungwirth2012,
  manchon08} has been recently shown to be an interesting and useful
addition in the field of spintronics. This subject capitalizes on
adjoining a strong SOI normal metal next to a thin magnetic layer
\cite{thiaville12}. The SOI converts a charge current, $J_c$, to a
spin current, $J_s$, with an efficiency parametrized by
${\Theta_\text{SH}}$, the spin Hall angle \cite{dyakonov71,
  hirsch99}. Recently, it was demonstrated experimentally that the
spin current produced in this way can switch the magnetization in a
dot \cite{miron11,liu12} or can partially compensate the damping
\cite{ando08a,liu11,demidov11d}, allowing the lifetime of propagating
spin-waves \cite{an14} to be increased beyond their natural decay
time, $\tau$. These two effects open potential applications in storage
devices and in microwave signal processing.

The effect is based on the fact that the spin current $J_s$ exerts a
torque on the magnetization, corresponding to an effective damping
$\Gamma_s = \gamma J_s / (t_\text{FM} M_s)$, where $t_\text{FM}$ is
the thickness of the magnetic layer, $M_s$ its spontaneous
magnetization, and $\gamma$ the gyromagnetic ratio. In the case of
metallic ferromagnets \cite{demidov12,liu12b,liu13}, it was
established that $\Gamma_s$ can fully compensate the natural damping
$1/\tau$ at a critical spin current $J_s^*$, which determines the
onset of auto-oscillation of the magnetization:
\begin{equation}
  J_s^*=- \, \dfrac 1 \tau \, \dfrac { t_\text{FM} M_s} \gamma \, .
\label{eq:Js}
\end{equation}
An important benefit of the SOI is that $J_c$ and $J_s$ are linked
through a cross-product, allowing a charge current flowing in-plane to
produce a spin current flowing out-of-plane. Hence it enables the
transfer of spin angular momentum to non-metallic materials and in
particular to insulating oxides, which offer improved performance
compared to their metallic counterparts. Among all oxides, Yttrium
Iron Garnet (YIG) holds a special place for having the lowest known
spin-wave (SW) damping factor. In 2010, Kajiwara \textit{et al.}
reported on the efficient transmission of spin current through the
YIG{\textbar}Pt interface \cite{kajiwara10}. It was shown that $J_s$
produced by the excitation of ferromagnetic resonance (FMR) in YIG can
cross the YIG{\textbar}Pt interface and be converted into $J_c$ in Pt
through the inverse spin Hall effect (ISHE). This finding was
reproduced in numerous experimental works \cite{sandweg10,
  kurebayashi11, vilela-leao11, chumak12, hahn13, kelly13, wang13}. In
the same paper, the reciprocal effect was also reported as $J_s$
produced in Pt by the direct spin Hall effect (SHE) could be
transferred to the 1.3~$\mu$m thick YIG, resulting in damping
compensation. However, attempts to directly measure the expected
change of the resonance linewidth of YIG as a function of the dc
current have so far failed \cite{hahn13,kelly13} \footnote{Indirect
  observation through small changes of the SW amplitude \cite{wang11,
    padron-hernandez11} cannot be conclusive due to the sensitivity to
  thermal effects.}. This is raising fundamental questions about the
reciprocity of the spin transparency, $T$, of the interface between a
metal and a magnetic insulator. This coefficient enters in the ratio
between $J_c$ in Pt and $J_s$ in YIG through:
\begin{equation}
  {J_s} = T\, {\Theta_\text{SH}} \, \dfrac{\hbar}{2e}\, {J_c}\,\,,
\label{eq:Jc}
\end{equation}
where $e$ is the electron charge and $\hbar$ the reduced Planck
constant. $T$ depends on the transport characteristics of the normal
metal as well as on the spin-mixing conductance $G_{\uparrow
  \downarrow}$, which parametrizes the scattering of the spin angular
momentum at the YIG{\textbar}Pt interface \cite{chen13}.

At the heart of this debate lies the exact value of the threshold
current. The lack of visible effects reported in
Refs.\cite{hahn13,kelly13}, although inconsistent with
\cite{kajiwara10}, is coherent with the estimation of the threshold
current of $10^{11-12}$~A.m$^{-2}$ using Eqs.(\ref{eq:Js}) and
(\ref{eq:Jc}) and typical parameters for the materials
\cite{xiao12}. This theoretical current density is at least one order
of magnitude larger than the maximum $J_c$ that could be injected in
the Pt so far. Importantly, the previous reported experiments were
performed on large (millimeter sized) structures, where many nearly
degenerate SW modes compete for feeding from the same dc source of
angular momentum, a phenomenon that could become self-limiting and
prevent the onset of auto-oscillations \cite{demidov11d}. To isolate a
single candidate mode, we have recently reduced the lateral dimensions
of the YIG pattern, as quantization results in increased frequency gaps
between the dynamical modes \cite{hahn14}. This requires to grow very
thin films of high quality YIG \cite{sun12, althammer13, wang13,
  pirro14}. Benefiting from our progress in the epitaxial growth of
YIG films by pulsed laser deposition (PLD) \cite{kelly13}, we propose
to study the FMR linewidth as a function of the dc current in a
micron-size YIG{\textbar}Pt disk.

\begin{figure}
  \includegraphics[width=8.5cm]{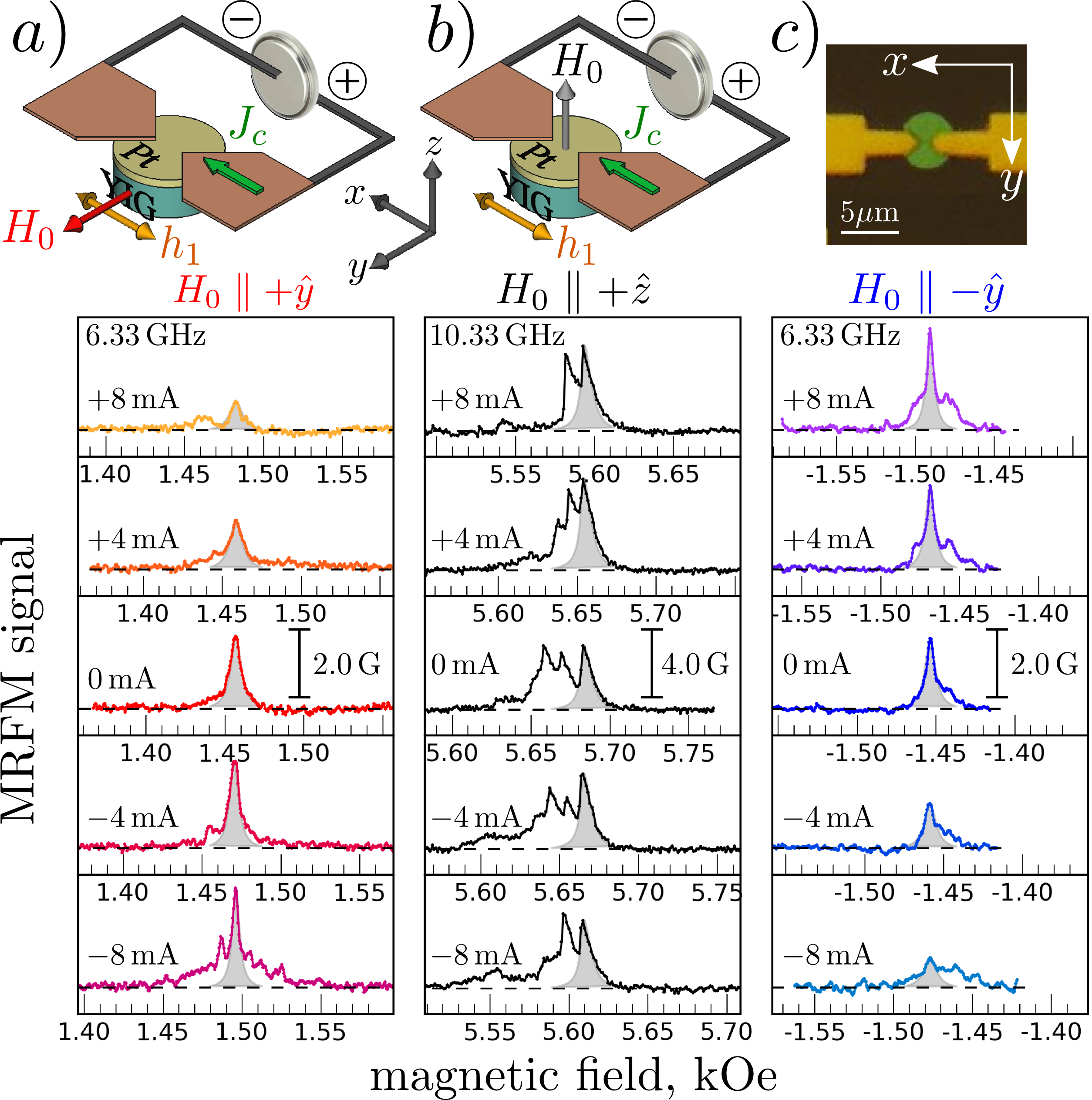}
  \caption{(Color online) MRFM spectra of the YIG{\textbar}Pt
    microdisk as a function of current for different field
    orientations: a) $H_0 \parallel +\hat y$ at $f=6.33$~GHz (red tone); b)
    $H_0 \parallel +\hat z$ at $f=10.33$~GHz (black); c)
    $H_0 \parallel -\hat y$ at $f=6.33$~GHz (blue tone). The highest
    amplitude mode is used for linewidth analysis (shaded area).
    Field axes are shifted so as to align the peaks
    vertically. In-plane and out-of-plane field orientations are
    sketched above. The top right frame is a microscopy image of the
    sample.}
  \label{FIG1}
\end{figure}

FIG.\ref{FIG1} shows a schematic of the experimental setup. A
YIG{\textbar}Pt disk of $5~\mu$m in diameter is connected to two Au
contact electrodes (see the microscopy image) across which a positive
voltage generates a current flow $J_c$ along the $+\hat
x$-direction. The microdisk is patterned out of a 20~nm thick
epitaxial YIG film with a 7~nm thick Pt layer sputtered on top. The
YIG and Pt layers have been fully characterized in previous studies
\cite{kelly13, rojas14}. Their characteristics are reported in Table
\ref{tab:mat}.

\begin{table}
  \caption{Transport and magnetic properties of the Pt and bare
    YIG layers, respectively from Ref.\cite{rojas14} and
    Ref.\cite{kelly13}.}
  \begin{ruledtabular}
    \begin{tabular}{c | c c c c }
      Pt & $t_\text{Pt}$ (nm) & $\sigma$ ($\Omega^{-1}$.m$^{-1}$) &
      $\lambda_\text{SD}$ (nm) & $\Theta_\text{SH}$ \\ \hline 
      & 7 & $5.8 \cdot 10^6$ & 3.5 & 0.056\\ \hline \hline 
      YIG & $t_\text{YIG}$ (nm) & $4 \pi M_s$ (G) & $\gamma$ (rad.s$^{-1}$.G$^{-1}$) & {$\alpha_0$} \\ \hline
      & 20 & $2.1 \cdot 10^3$ & $1.79 \cdot 10^{7}$ & {$2.3 \cdot 10^{-4}$} 
 \end{tabular}
\end{ruledtabular}\label{tab:mat}
\end{table}

The sample is mounted inside a room temperature magnetic resonance
force microscope (MRFM) which detects the SW absorption spectrum
mechanically \cite{zhang96,klein08,chia12}. The excitation is provided
by a stripline (not shown in the sketches of FIG.\ref{FIG1})
generating a linearly polarized microwave field $h_1$ along the $\hat
x$-direction. The detection is based on monitoring the deflection of a
mechanical cantilever with a magnetic Fe particle affixed to its tip,
coupled dipolarly to the sample. The FMR spectrum is obtained by
recording the vibration amplitude of the cantilever while scanning the
external bias magnetic field, $H_{0}$, at constant microwave
excitation frequency, $f=\omega/(2 \pi)$ \footnote{When FMR conditions
  are met, the precession angle increases and in consequence the
  static magnetization (projection along the precession axis)
  decreases, hereby exciting mechanically the cantilever.}. The MRFM
is placed between the poles of an electromagnet, generating a uniform
magnetic field, $H_0$, which can be set along $\hat{y}$ or $\hat{z}$
(\textit{i.e.}, perpendicularly to both $h_1$ and $J_c$).

We start by measuring the effect of a dc current, $I_\text{dc}$, on
the FMR spectra when the disk is magnetized in-plane by a magnetic
field along the $+\hat y$-direction (positive field). The spectra
recorded at $f=6.33$~GHz are shown in FIG.\ref{FIG1}a in red
tones. The middle row shows the absorption at zero current. The MRFM
signal corresponds to a variation of the static magnetization of about
2\,G, \textit{i.e.}, a precession cone of 2.5$^\circ$. As the the
electrical current is varied, we observe very clearly a change of the
linewidth. At negative current, the linewidth decreases, to reach
about half the initial value at $I_\text{dc}=-8$~mA. This decrease is
strong enough so that the individual modes can be resolved
spectroscopically within the main peak. Concomitantly the amplitude of
the MRFM signal increases. The opposite behavior is observed when the
current polarity is reversed. At positive current, the linewidth
increases to reach about twice the initial value at
$I_\text{dc}=+8$~mA, and the amplitude of the signal decreases.

$I_\text{dc}=\pm12$~mA is the maximum current that we have injected in
our sample to avoid irreversible effects.  We estimate from the Pt
resistance, the sample temperature to be $90^\circ$~C at the maximum
current. This Joule heating reduces $4\pi M_s$ at a rate of $4.8$~G/K,
which results in an even shift of the resonance field towards higher
field \footnote{Actually, the resonance peak shifts in field due to
  both the linear ($\approx 1.5$~Oe/mA) and quadratic ($\approx
  0.5$~Oe/mA$^2$) contributions in current of Oersted field and Joule
  heating, respectively.}.

In FIG.\ref{FIG1}b, we show the FMR spectra at $f=10.33$~GHz in the
perpendicular geometry, \textit{i.e.}, $H_0$ is along $\hat z$. In
contrast to the previous case, the linewidth does not change with
current. This is expected as no net spin transfer torque is exerted by
the spin current on the precessing magnetization in this
configuration. Note that due to Joule heating, the spectrum now shifts
towards lower field due to the decrease of $M_s$ as the current
increases.

We now come back to the in-plane geometry, but this time, the magnetic
field is reversed compared to FIG.\ref{FIG1}a, \textit{i.e.}, applied
along $-\hat y$ (negative field). The corresponding spectra are
presented in FIG.\ref{FIG1}c using blue tones. As expected for the
symmetry of the SHE, the observed behavior is inverted with respect to
FIG.\ref{FIG1}a: a positive (negative) current now reduces (broadens)
the linewidth.

\begin{figure}
  \includegraphics[width=7.cm]{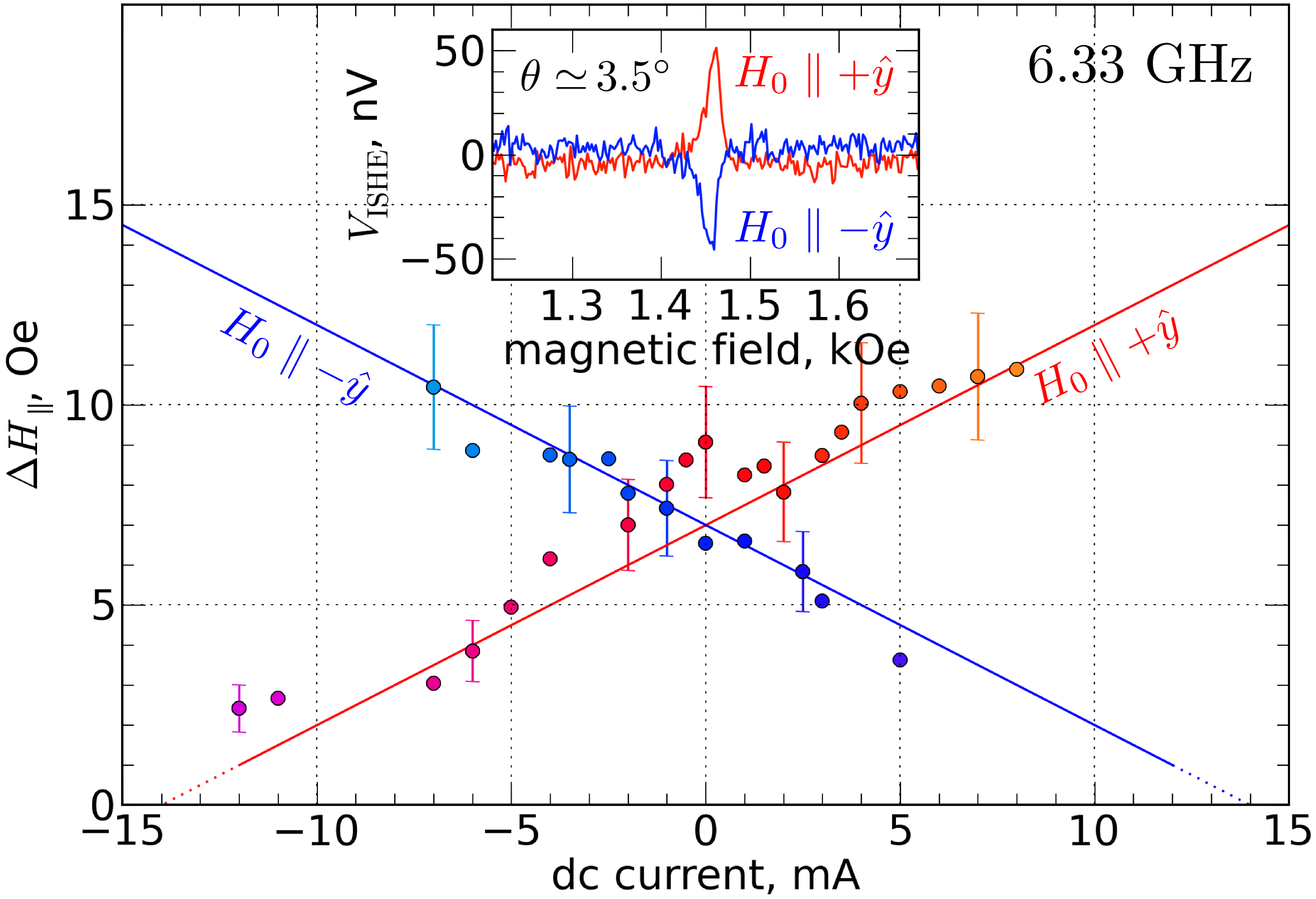}
  \caption{(Color online) Variation of the full linewidth $\Delta
    H_\parallel$ measured at 6.33~GHz as a function of $I_\text{dc}$
    for $H_0 \parallel +\hat y$ (red) and $H_0 \parallel -\hat y$
    (blue). Inset: detection of $V_\text{ISHE}$ as a function of $H_0$
    at $f=6.33$~GHz and $I_\text{dc}=0$.}
  \label{FIG2}
\end{figure}

We report in FIG.\ref{FIG2} the values of $\Delta H_\parallel$, the
full linewidth measured in the in-plane geometry, as a function of
current. The data points follow approximately a straight line, whose
slope $\pm0.5$~Oe/mA reverses with the direction of $H_0$ along $\pm
\hat{y}$ and whose intercept with the abscissa axis occurs at
$I^*_\text{6.33 GHz}=\mp14$~mA. Moreover, we emphasize that the
variation of linewidth covers about a factor five on the full range of
current explored.

The inset of FIG.\ref{FIG2} shows the inverse spin Hall voltage
$V_\text{ISHE}$ measured at $I_\text{dc}=0$~mA and $f=6.33$~GHz. This
voltage results from the spin current produced by spin pumping from
YIG to Pt and its subsequent conversion into charge current by ISHE
\cite{kajiwara10}. Its sign changes with the direction of the bias
magnetic field, as shown by the blue and red $V_\text{ISHE}$
spectra. This observation confirms that a spin current can flow from
YIG to Pt and that damping reduction occurs for a current polarity
corresponding to a negative product of $V_\text{ISHE}$ and
$I_\text{dc}$ .

\begin{figure}
  \includegraphics[width=8.5cm]{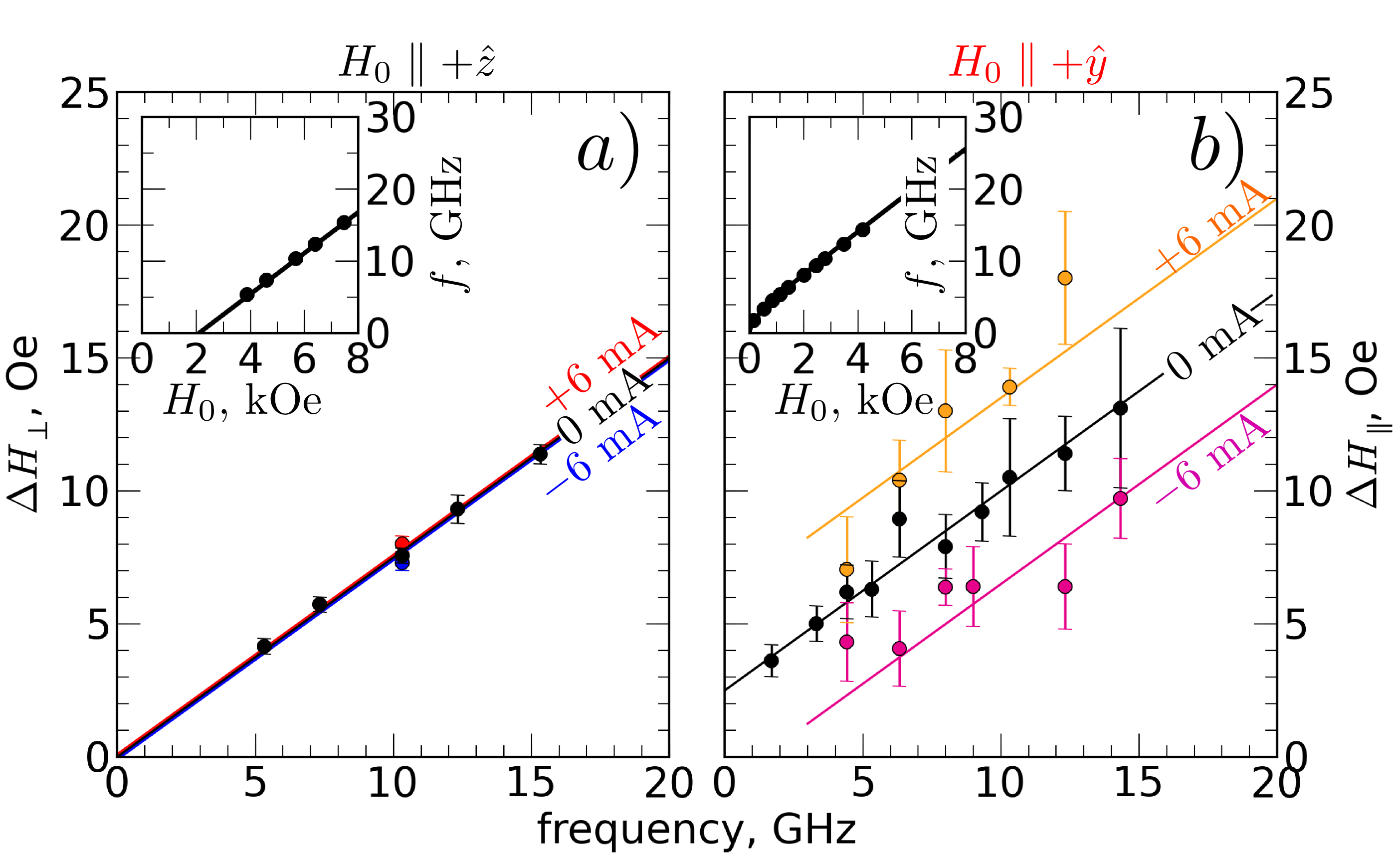}
  \caption{(Color online) Frequency dependence of the linewidth for
    three values of the dc current (0, $\pm$6~mA) a) in the
    perpendicular geometry and b) in the parallel geometry. Insets
    show the corresponding dispersion relations $f(H_0)$.}
  \label{FIG3}
\end{figure}

To gain more insight into these results, we now analyze the frequency
dependence of the full linewidth at half maximum for three values of
dc current (0, $\pm 6$~mA) for both the out-of-plane and in-plane
geometries. We start with the out-of-plane data, plotted in
FIG.\ref{FIG3}a. The dispersion relation displayed in the inset
follows the Kittel law, $\omega = \gamma (H_0 - 4 \pi N_\text{eff}
M_s)$, where $N_\text{eff}$ is an effective demagnetizing factor close
to 1 \cite{kakazei04,naletov11}. The linewidth $\Delta H_\perp$
increases linearly with frequency along a line that intercepts the
origin, a signature that the resonance is homogeneously broadened
\cite{hahn14}. In this geometry, the Gilbert damping coefficient is
simply $\alpha = \gamma \Delta H_\perp/ (2 \omega)= 1.1 \cdot 10^{-3}$
and the reaxation time $\tau=1/(\alpha \omega)$. We also report
on this figure the fact that at 10.33~GHz, $\Delta H_\perp =$ 7~Oe is
independent of the current (see FIG.\ref{FIG1}b).

The damping found in our YIG{\textbar}Pt microdisk is significantly
larger than the one measured in the bare YIG film $\alpha_0
=2.3\cdot10^{-4}$ (cf. Table \ref{tab:mat}). This difference is due to
the spin pumping effect, and enables to determine the spin-mixing
conductance of our YIG{\textbar}Pt interface through
\cite{tserkovnyak05,heinrich11}:
\begin{equation}
  \alpha=\alpha_0+\frac{\gamma \hbar}{4 \pi M_st_\text{YIG}}\frac{G_{\uparrow \downarrow}}{G_0} \, ,
  \label{eq:alpha}
\end{equation}
where $G_0=2e^2/h$ is the quantum of conductance. The measured
increase of almost $9\cdot10^{-4}$ for the damping corresponds to
$G_{\uparrow \downarrow}=1.5 \times 10^{14} \Omega^{-1}\text{m}^{-2}$,
in agreement with a previous determination made on similar
YIG{\textbar}Pt nanodisks \cite{hahn14}. This value allows us to
estimate the spin transparency of our interface \cite{chen13},
$T=G_{\uparrow \downarrow}/(G_{\uparrow
  \downarrow}\coth{\left({t_\text{Pt}}/{\lambda_\text{sd}}\right)}+
{\sigma}/({2\lambda_\text{sd}}))\simeq0.15$, where $\sigma$ is the Pt
conductivity and $\lambda_\text{sd}$ its spin-diffusion
length. Moreover, the spin-mixing conductance can be used to analyze
quantitatively the dc ISHE voltage produced at resonance
\cite{mosendz10,castel12a,hahn13}. Using the parameters of Table
\ref{tab:mat} and the value of $G_{\uparrow \downarrow}$, we find that
the 50~nV voltage measured in the inset of FIG.\ref{FIG2} is produced
by an angle of precession $\theta \simeq 3.5^\circ$, which lies in the
expected range.

We now turn to the in-plane data, presented in FIG.\ref{FIG3}b. The
dispersion relation plotted in the inset follows the Kittel law
$\omega = \gamma \sqrt{H_0 (H_0 + 4 \pi N_\text{eff} M_s)}$. In this
case, $1/\tau = \alpha \, (\partial \omega/\partial H_0)\,(\omega /
\gamma)$. For $I_\text{dc}=0$~mA the slope of the linewidth
vs. frequency is exactly the same as that in the perpendicular direction
$\alpha = 1.1 \cdot 10^{-3}$. For this geometry, however, the line
does not intercept the origin, indicating a finite amount of
inhomogeneous broadening $\Delta H_0 = 2.5$~Oe, \textit{i.e}, the
presence of several modes within the resonance line. Setting
$I_\text{dc}$ to $\pm 6$~mA shifts $\Delta H_\parallel$ by $\pm 3$~Oe
independently of the frequency, which is consistent with the rate of
0.5~Oe/mA reported at 6.33~GHz in FIG.\ref{FIG2}. In fact, in the
presence of the effective damping $\Gamma_s$, the linewidth of the
resonance line varies as
\begin{equation}
  \Delta H_\parallel = \Delta H_0 + 2 \alpha \dfrac \omega \gamma + 2 \dfrac {J_s}{M_s t_\text{YIG}} \, .
\label{eq:DH}
\end{equation}
This expression is valid when $(\partial \omega/\partial H_0) \simeq
\gamma$, \textit{i.e.}, at large enough field or frequency (see inset
of FIG.\ref{FIG3}b). It describes appropriately the experimental data on
the whole frequency range measured.

\begin{figure}
  \includegraphics[width=8.5cm]{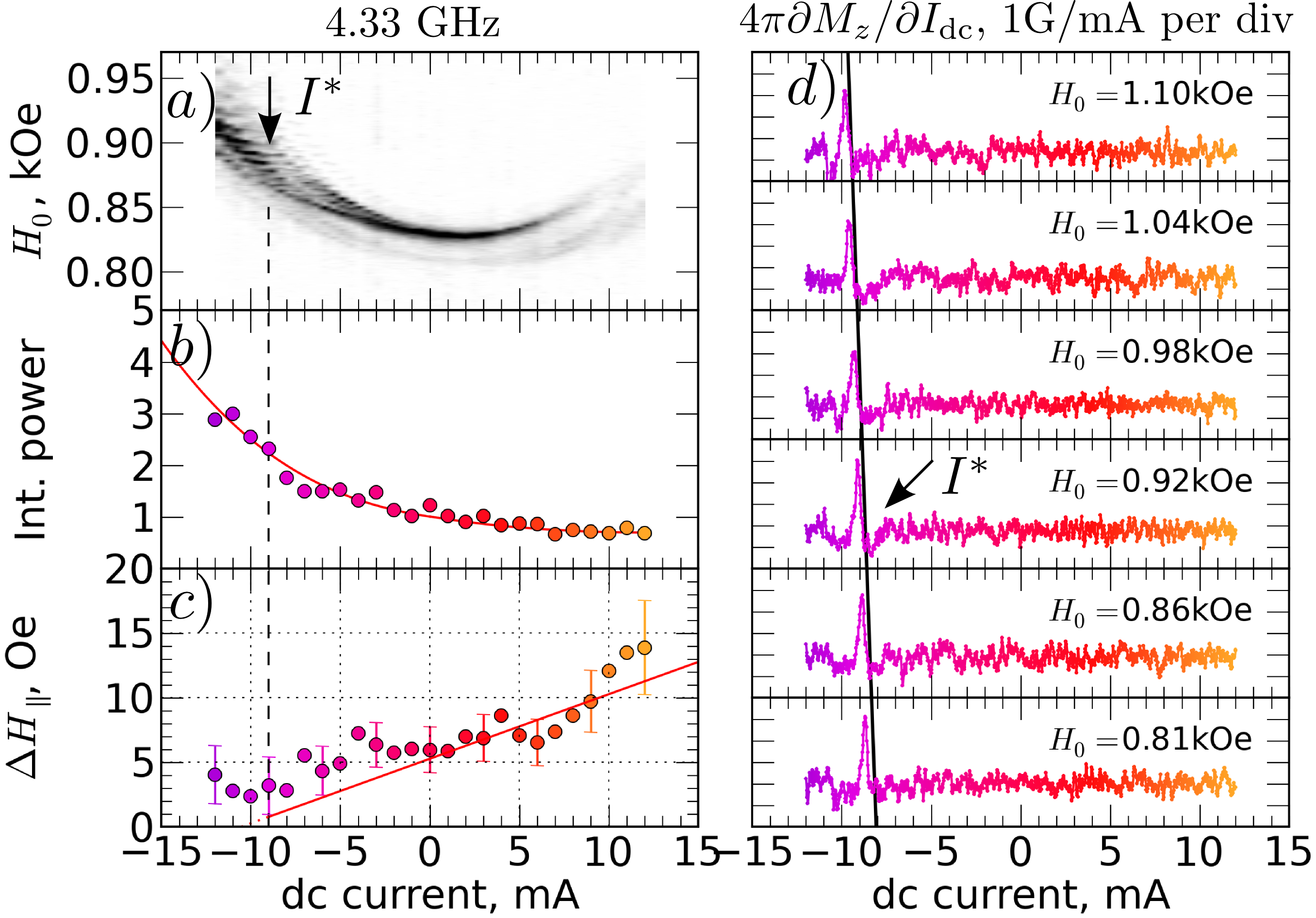}
  \caption{(Color online) a) Density plot of the MRFM spectra at
    4.33~GHz vs. field and current $I_\text{dc} \in [-12,+12]$~mA. The
    color scale represents $4\pi\Delta M_z$ (white: 0~G, black:
    1.5~G). b) Evolution of integrated power vs. $I_\text{dc}$. c)
    Dependence of linewidth on $I_\text{dc}$. d) Differential
    measurements of $M_z$ ($I_\text{dc}$ modulated by
    0.15~mA$_\text{pp}$, no rf excitation) vs. $I_\text{dc}$ at six
    different values of the in-plane magnetic field.}
  \label{FIG4}
\end{figure}

In order to investigate the autonomous dynamics of the YIG layer and
exceed the compensation current, $I^*$, we now perform measurements at
lower excitation frequency, where the threshold current is estimated
below 12~mA. In FIG.\ref{FIG4}a, we present a density plot of the MRFM
spectra acquired at 4.33~GHz as a function of the in-plane magnetic
field and $I_\text{dc}$ through the Pt. The measured signal is clearly
asymmetric in $I_\text{dc}$. At positive current, it broadens and its
amplitude decreases, almost disappearing above $+ 8$~mA, whereas at
negative current, it becomes narrower and the amplitude is maximal at
$I_\text{dc}<-10$~mA.

The power integrated over the full field range normalized by its value
at 0 mA and the linewidth variation vs. $I_\text{dc}$ are plotted in
FIGs.\ref{FIG4}b and \ref{FIG4}c, respectively. The normalized
integrated power varies by a factor of five from $+12$~mA to $-12$~mA
following an inverse law on $I_\text{dc}$ (see continuous line), which
is consistent with the spin transfer effect
\cite{demidov11d,hamadeh12}. The linewidth varies roughly linearly
with $I_\text{dc}$: it increases from 6~Oe at 0~mA up to 14~Oe at
$+12$~mA and it reaches a minimum value close to 2~Oe between $-8$ and
$-11$~mA. It is interesting to note that this happens in a region of
the density plot where the evolution of the signal displays some kind
of discontinuity, with the appearance of several high amplitude peaks
in the spectrum (see arrow in FIG.\ref{FIG4}a). We tentatively ascribe
this feature to the onset of auto-oscillations in the YIG layer,
namely, one or several dynamical modes have their relaxation
compensated by the injected spin current and are destabilized
\cite{kajiwara10}.

To confirm this hypothesis, we present in FIG.\ref{FIG4}d results of
an experiment where no rf excitation is applied to the system. Here,
the dc current is modulated at the MRFM cantilever frequency by
$\delta I=0.15$~mA$_\text{pp}$ and the induced $\delta M_z$ is probed
as a function of $I_\text{dc}$. This experiment thus provides a
differential measurement $\partial M_z/\partial I_\text{dc}$ of the
magnetization (in analogy with $dV/dI$ measurements in transport
experiments). At $H_0=0.92$~kOe, a peak in $\partial M_z/\partial
I_\text{dc}$ is measured around $-9$~mA. It corresponds to a variation
of $4\pi\delta M_z\simeq0.5$~G, \textit{i.e.}, a change of the angle
of precession by $1.3^\circ$ induced by the modulation of
current. Moreover, this narrow peak observed in $\partial M_z/\partial
I_\text{dc}$ shifts linearly in dc current with the applied magnetic
field, from $-8$~mA at 0.81~kOe to $-10$~mA at 1.1~kOe (see the
continuous straight line in FIG.\ref{FIG4}d), in agreement with the
expected behavior of the threshold current Eq.(\ref{eq:Js}).

Hence, FIG.\ref{FIG4} presents a set of data consistent with the
determination of a critical current of $I^*=-9$~mA at $H_0=0.92$~kOe,
corresponding to $J_c^*\simeq3\cdot10^{11}$~A.m$^{-2}$, in agreement
with the value of $2\cdot10^{11}$~A.m$^{-2}$ expected from
Eqs.(\ref{eq:Js}) and (\ref{eq:Jc}) and the parameters of our
system. Nevertheless, the destabilization of dynamical modes is rather
small, as the jump of resonance field at $I^*$ (due to reduction of
the magnetization) does not exceed the linewidth. We suspect that in
our YIG$|$Pt microdisk, the splitting of modes is not sufficient to
prevent nonlinear interactions that limit the amplitude of
auto-oscillations \cite{demidov11d}. In order to favor larger
auto-oscillation amplitudes, YIG structures that are even more
confined laterally (below 1 $\mu$m) should be used \cite{hahn14}, or
one should excite a bullet mode \cite{demidov12}.

In conclusion, we have demonstrated that it is possible to control
electronically the SW damping in a YIG microdisk. Extending this
result to one-dimensional SW guide \cite{duan14} will offer great
prospect in the emerging field of magnonics
\cite{kruglyak10a,serga10}, whose aim is to investigate the
manipulation of SW and their quanta -- magnons -- with the benefice of
combining ultra-low energy consumption and compactness. To improve the
magnonic paradigm, a solution will be to actively compensate damping
in the YIG magnetic insulator by SW amplification through stimulated
emission generated by a charge current in the adjacent metallic layer
with strong SOI.

\begin{acknowledgments}
  This research was supported by the French Grants Trinidad (ASTRID
  2012 program), by the RTRA Triangle de la Physique grant Spinoscopy,
  and by the Deutsche Forschungsgemeinschaft. We acknowledge
  C. Deranlot, E. Jacquet, and R. Lebourgeois for their contribution
  to the growth of the sample and A. Fert for fruitful discussion.
\end{acknowledgments}


%

\end{document}